\documentclass[12pt, letterpaper]{article}
\usepackage[top=1in, bottom=1.5in, left=1in, right=1in]{geometry}
\usepackage[utf8]{inputenc}
\usepackage{booktabs}
\usepackage{hyperref}
\usepackage{graphicx}
\usepackage{gensymb}
\usepackage{natbib}
\usepackage[font=footnotesize]{caption}
\usepackage[strict]{changepage}
\newcommand{\ttt}[1]{\texttt{#1}}

\title{Mapping the Periphery and Variability of the Magellanic Clouds}
\author{Knut Olsen, Paula Szkody, Maria-Rosa Cioni, Marcella Di Criscienzo,\\
Ilaria Musella, Vincenzo Ripepi, Francesco Borsa, Marcella Marconi,\\
L\'eo Girardi, Giada Pastorelli, Michele Trabucchi, Paolo Ventura,\\
Marc Moniez \\
with the support of the LSST SMWLV and TVS Collaborations}
\date{November 2018}

\begin{document}

\maketitle

\begin{abstract}
We propose a combination of a modified Wide-Fast-Deep survey, a mini-survey of the South Celestial Pole, and a Deep Drilling-style survey to produce a 3-D map of the Magellanic System and to provide a detailed census of the transient and variable populations in the Clouds.  We support modifying the Wide-Fast-Deep survey to cover the declination range $-72.25\deg<{\rm Dec}<12.4\deg$ and the Galactic latitude range $|b|>15\deg$, as proposed in a separate white paper. We additionally propose a mini-survey covering the 950$\deg^2$ with ${\rm Dec} < -72.25$ in $ugriz$ to the standard LSST single-exposure depth and with 40 visits per filter per field.  Finally, we propose a mini-survey covering $\sim100 \deg^2$ of the main bodies of the Clouds with twelve total pointings, 2000 total visits per field, and shorter exposure time.
\end{abstract}

\section{White Paper Information}
\begin{enumerate}
    \item {\bf Contact information:} Knut Olsen, kolsen@noao.edu;\\
    Paula Szkody, szkody@astro.washington.edu
    \item {\bf Science Category:} Structure and Stellar Content of the Milky Way, Exploring the Transient/Variable Universe
    \item {\bf Survey Type Category:} Wide-Fast-Deep, Mini survey, Deep Drilling field
    \item {\bf Observing Strategy Category:} an integrated program with science that hinges on the combination of pointing and detailed 
	observing strategy
\end{enumerate}


\clearpage

\section{Scientific Motivation}
The Magellanic Clouds (MCs) have always had outsized importance for astrophysics.  They are critical steps in the cosmological distance ladder, they are a binary galaxy system with a unique interaction history, and they are laboratories for studying all manner of astrophysical phenomena, giving us front-row seats to observe the life cycles of galaxies.  They are often used as jumping-off points for investigations of much larger scope and scale; examples are the searches for extragalactic supernovae prompted by the explosion of SN1987A and the dark matter searches through the technique of gravitational microlensing. 

LSST has a unique role to play in surveys of the Clouds.  First, its large $A\Omega$ will probe the thousands of square degrees that comprise the extended periphery of the Magellanic Clouds with unprecedented completeness and depth, allowing us to detect and map their extended disks, stellar halos, and debris from interactions with high fidelity.  Second, the ability of LSST to map the entire main bodies in only a few pointings will allow us to identify and classify their extensive variable source populations with high time and areal coverage, discovering, for example, extragalactic planets, rare variables and transients, and light echoes from explosive events that occurred thousands of years ago.  History has shown that discoveries made in the Clouds often have highly interconnected dependencies, such that the topics above will likely feed on each other in unexpected ways.  Surveys such as the Dark Energy Survey (DES) and Gaia have delivered surprising new results but also shown how much remains to be discovered in the environs of the Clouds.
%
\vspace{0.1in}

\begin{figure}[ht!]
     \begin{adjustwidth}{-3em}{-3em}
     \centering
    \includegraphics[width=1.0\columnwidth,trim=0.0in 0.0in 0.0in 0.0in]{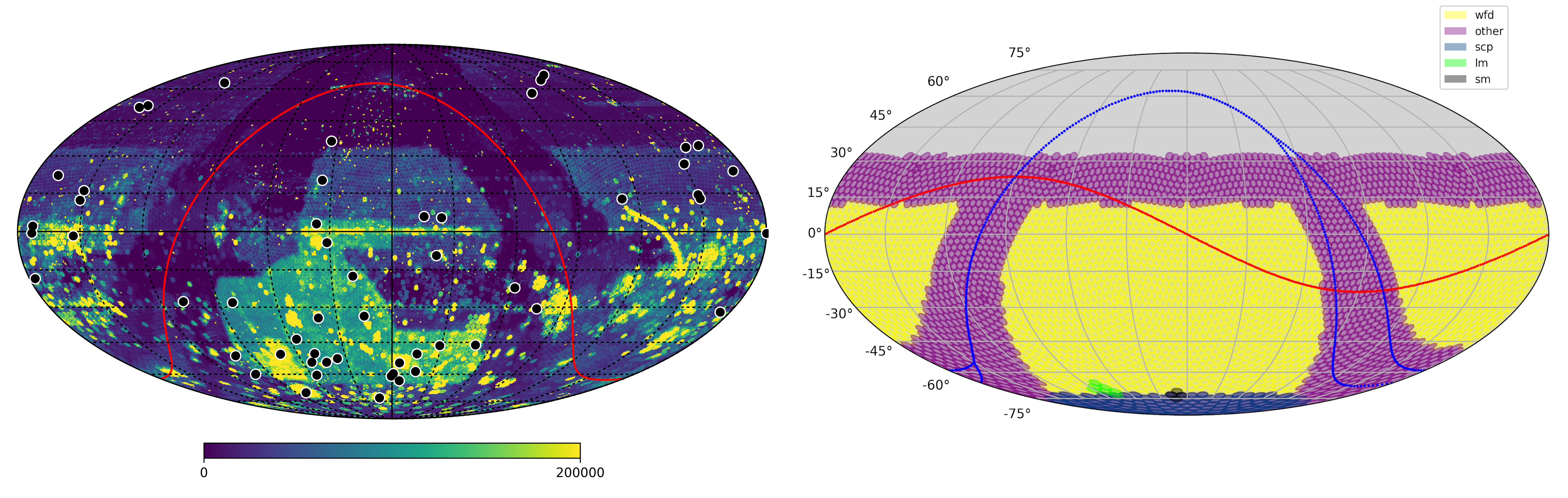}
    \caption{
    {\it Left:} The dwarf companions of the Milky Way and Magellanic Clouds (black dots), projected against the NOAO DECam+Mosaic map of total archive exposure time.  Deep surveys such as DES have added many new dwarf galaxy detections, as well as stellar streams. The concentration of new dwarfs in the DES footprint is  aided by the depth and uniformity of DES, but may also reflect an association with the Magellanic Clouds.  The red line marks the Galactic Plane. {\it Right:} Our proposed modification of the LSST survey footprint includes WFD (yellow area), the SCP (dark blue area), and the main bodies of the Magellanic Clouds (green and black areas).  The modification of the footprint maximizes our ability to map the structure of the Milky Way outside the Plane and the Magellanic System. The red line marks the Ecliptic Plane.
    }
    \label{fig:mcdwarfs}
    \end{adjustwidth}
\end{figure}

\begin{figure}[hbt!]
     \begin{adjustwidth}{-3em}{-3em}
    \centering
    \includegraphics[width=0.6\columnwidth,trim=0.0in 0.0in 0.0in 0.0in]{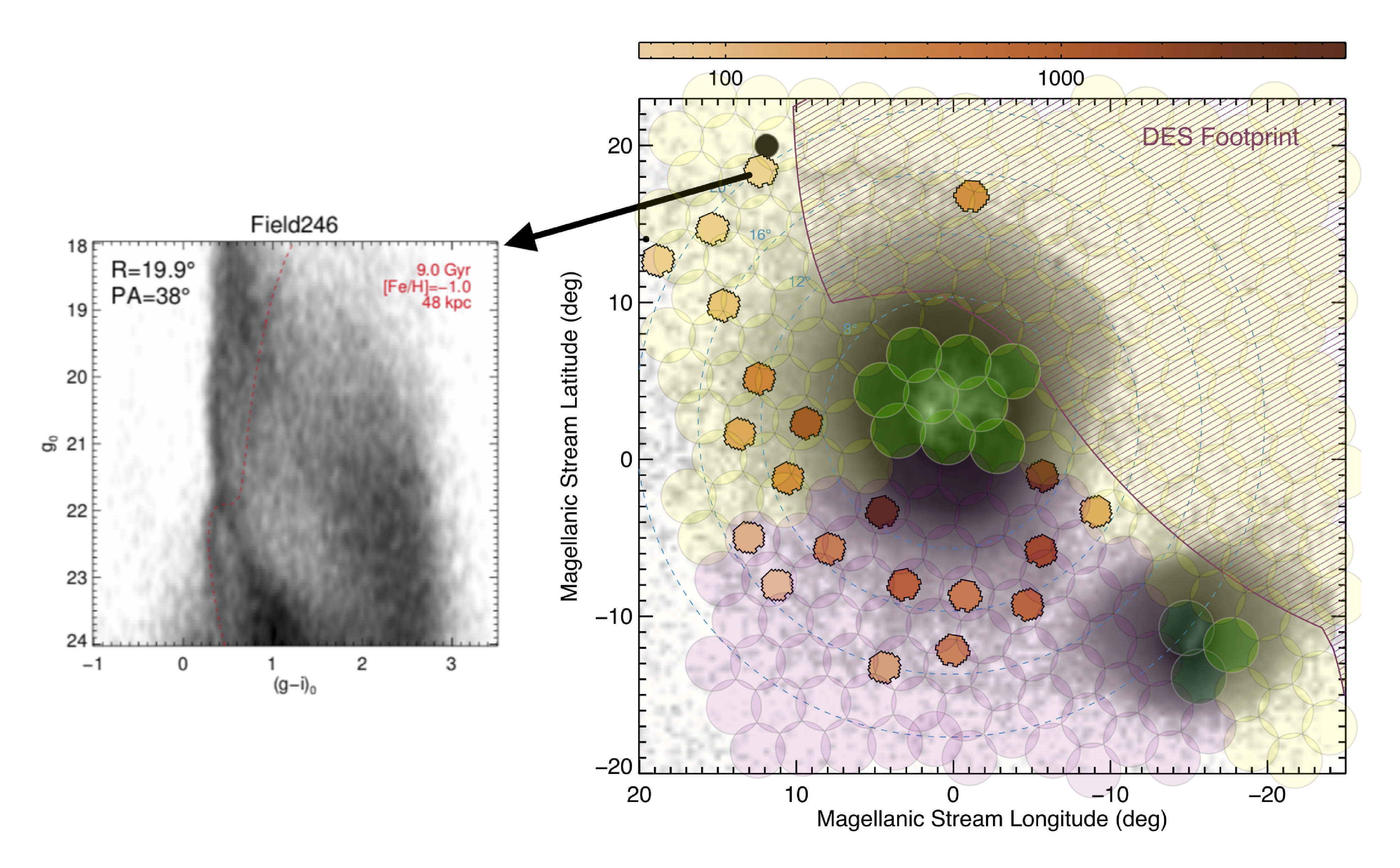}
    \caption{
    Stellar structure in the periphery of the MCs. In the map on the right, RGB stars observed with Gaia (grayscale; \citep{Belokurov2019}) show that substructure from the Magellanic Clouds extends over a vast area.  Overlaid are fields observed
    by the DECam-based SMASH survey (shaded hexagons), color-coded by the density of MSTO stars \citep{Nidever2018}, as well as our proposed LSST pointings (WFD: yellow; SCP: purple; MC main body fields: green) and the DES footprint (hatched area).  MSTO stars are $\sim$50$\times$ more numerous than RGB stars, making them powerful tracers of low surface brightness structure.  The panel on the left shows a CMD from one SMASH field, where MSTO stars from the LMC are seen projected against the Galactic foreground. In the WFD area, the LSST map will be $>$3 magnitudes deeper, while in the SCP area it will be 2 magnitudes deeper.  In the main body fields we will observe variability on all timescales, from 30 seconds to 10 years. }
    \label{fig:mcperiphery}
    \end{adjustwidth}
\end{figure}

\noindent\textbf{A 3-D Map of the Magellanic System}\\
Over the past decade, our view of the MCs has changed substantially.  We now have evidence that they fell into our Galaxy's halo only recently \citep{Kallivayalil2013}, that they have a satellite galaxy system of their own (\citep{Bechtol2015,Koposov2015}; Fig.\ref{fig:mcdwarfs}), and that their binary interaction has produced a field of stellar debris spreading over at least a 20 degree radius from their centers (\citep{Saha2010,Majewski2009,Nidever2018,Belokurov2019}; Fig.~\ref{fig:mcperiphery}).  These discoveries increase the relevance of the MCs to the study of the Milky Way and galaxy formation and evolution generally, as they are probes of the effects of interactions on galaxy structure and star formation, of galaxy host properties on their satellite populations, and of group infall on the galaxies themselves and on the Milky Way system.

LSST has the opportunity to provide the ultimate map of the extent and structure of the Magellanic system.  Our current maps are based on complete surveys of relatively rare red giant tracers or on partially filled surveys of the much more numerous main sequence turnoff (MSTO) stars, down to depths of $r\sim$24.  With photometry from image stacks as faint as $r\sim27$, LSST will be able to detect extremely low surface brightness structure and dwarf companions of the MCs.  From single-epoch photometry curves, LSST will detect both RR Lyrae and $\delta$ Scuti variables associated with this structure, giving us a 3D map of the MCs, their surroundings, and their satellite populations.  Moreover, LSST's multiband coverage will characterize the ages and metallicities of these stellar populations, for the clearest possible view of the Magellanic system and its relationship to the Milky Way.

\vspace{0.1in}
\noindent\textbf{A Detailed Census of Variables and Transients}\\
The main bodies of the MCs are hosts to a huge range of variables and transients, all with approximately the same distance, low foreground extinction, and low metallicity, making them obvious targets for collecting large statistical samples of known phenomena and for exploring the unknown. With LSST we will:
\begin{itemize}
    \item Obtain light curves and periods of the full range of variable objects, including regularly variable objects down to M$_V$=6.5, irregular variables, and eruptive and flaring objects, probing both the physics of the variable populations and the 3D structure of the Clouds;
    \item Observe light echoes from past supernovae and other explosive
    events, probing the physics of explosions through the study of their original light (e.g. \citep{Rest2005});
    \item Use interstellar scintillation from turbulent gas clouds along the line of sight to probe otherwise invisible baryonic matter \cite{Moniez2003};
    \item View giant planet transits of stars in the LMC, probing planet formation in a low metallicity galaxy \cite{Lund2015};
    \item Observe microlensing events from compact objects in the MCs and Galactic halo, with sensitivity to a wide spectrum of lens masses, including black holes
\end{itemize}
The science that can be done with this sample will be fundamental, will help to be build a complete and consistent picture of the Clouds as galaxies, and provide a touchstone for interpretation of all variables and transients discovered by LSST.  For example, eclipsing binaries will allow measurement of fundamental parameters of the stars and enable an independent age constraint for interpretation of star formation histories. Detection of outbursts in cataclysmic variables and AM CVn objects can delineate the correct space densities to compare with close-binary evolution models. The optical variability of super soft sources provides clues to the nature of the accreting objects as white dwarfs or black holes. Determining the flare characteristics of a large number of stars will advance our understanding of the differences between solar and stellar flares. Identifying pulsating white dwarfs provides information on the cooling curves of H, He white dwarfs, allowing a comparison of the IMF across a range of metallicities. Identification of $\delta$Scuti
and $\gamma$ Dor variables will provide a check on the unstable stellar evolution regions in 1.3-2.5 M$_{\odot}$ stars.

The timescale of variability of these objects span the full spectrum from 30 seconds to 10 years, and require the depth, area, and temporal coverage that only LSST can provide. Finally, the large number of observing opportunities provided by LSST will allow us to produce the highest resolution complete optical map of the Magellanic Clouds ever made.  Images taken of the main bodies of the MCs are ultimately confusion-limited.  A 0.4 arcsec resolution map of the MCs would be two magnitudes deeper than existing $\sim$1 arcsec maps.  Such a map would have great legacy value for pinpointing the progenitors of variable and transient objects, for studying the star formation history of the MCs, and as a reference for future surveys.


\clearpage


\clearpage
\vspace{.6in}

\section{Technical Description}


\subsection{High-level description}
We propose a combination of a modified Wide-Fast-Deep survey, a mini-survey of the South Celestial Pole, and a Deep Drilling-style survey to meet the goals of LSST Magellanic Clouds science.  In summary:
\begin{itemize}
\item We propose to adopt the main survey plan and WFD footprint advocated by the ``Big Sky" (Olsen et al.) white paper, where WFD is defined as the declination range $-72.25\deg<{\rm Dec}<12.4\deg$ and the Galactic latitude range $|b|>15\deg$.  This extends the WFD survey to the MCs and much of their periphery, with 825 total visits per field, and provides for an average of $\sim250$ visits per field for non-WFD parts of the sky {\em as part of the main survey}.  We have no strong constraints on the cadence for WFD, other than that it should allow for measurement of proper motions over the full 10-year baseline and should allow for the identification and period measurement of RR Lyrae stars.
\item We propose covering the 950$\deg^2$ (113 fields) around the SCP with ${\rm Dec} < -72.25\deg$ in $ugriz$ (no $y$) to the standard LSST single-exposure depth and with 40 visits per filter per field (for a total of 200 visits per field), yielding stacked depths of $(ugriz)=(25.6,26.8,26.4,25.8,24.9)$, or 2 mags deeper than single epoch depth.  The cadence constraints are the same as for WFD above.  Under our plan, these visits come as part of the main survey, in which all non-WFD fields receive a baseline of 250 visits per field.  By using only 200 visits per field for the SCP, we leave 50 visits$\times$113 fields = $\sim$6100 visits to use for the MC main bodies (see next item), {\em all within the main survey budget.}
\item We propose a deep drilling-style survey covering the main bodies of the Clouds with twelve total pointings and $\sim$2000 total visits per field to be completed in 10 years.  The cadence of these observations is described in detail below.  The visit budget for these fields breaks down as: 825 visits from WFD, (6100/12)=508 visits taken from SCP, and 667 visits per field from the 10\% of the survey time reserved for Deep Drilling and mini-surveys.  Given that 10\% of the visits is $\sim$260,000, this means that the MC main body deep-drilling survey uses only (667$\times$12)/260000=3\% of the time available for Deep Drilling and mini-surveys. 
\end{itemize}

\vspace{.3in}

\subsection{Footprint -- pointings, regions and/or constraints}
As described above, our proposed footprint includes:
\begin{itemize}
    \item WFD defined as $-72.25\deg<{\rm Dec}<12.4\deg$ and $|b|>15\deg$
    \item SCP region defined as ${\rm Dec} < -72.25$
    \item Twelve fields covering the MC main bodies, with coordinates in the table below
\end{itemize}

\begin{table}[h!]
\begin{tabular}{|l|l|l|l|}
\hline
fieldID & FOV & RA & DEC\\
\hline
\multicolumn{4}{|c|}{\textbf{ SMC}}\\
\hline
114 & 3.5 & 5.606889 & -73.293784\\
124 & 3.5 & 16.568751 & -72.896888\\
154 & 3.5 & 9.777261 & -70.757693\\
%
\hline
\multicolumn{4}{|c|}{\textbf{ LMC}}\\
\hline
138 & 3.5 & 72.000005 & -70.933132 \\
150 & 3.5 & 81.777206 & -70.757749 \\
166 & 3.5 & 91.216696 & -70.227751 \\
178 & 3.5 & 76.364237 & -68.470535\\
190 & 3.5 & 84.972540 & -68.155026\\
207 & 3.5 & 93.230817 & -67.520318\\
226 & 3.5 & 71.999949 & -66.069988\\
236 & 3.5 & 79.825808 & -65.928086\\
238 & 3.5 & 87.479642 & -65.499818\\
%
\hline
\end{tabular}
\end{table}

\subsection{Image quality}
\begin{figure}[h!]
\centering
\includegraphics[width=0.44\textwidth]{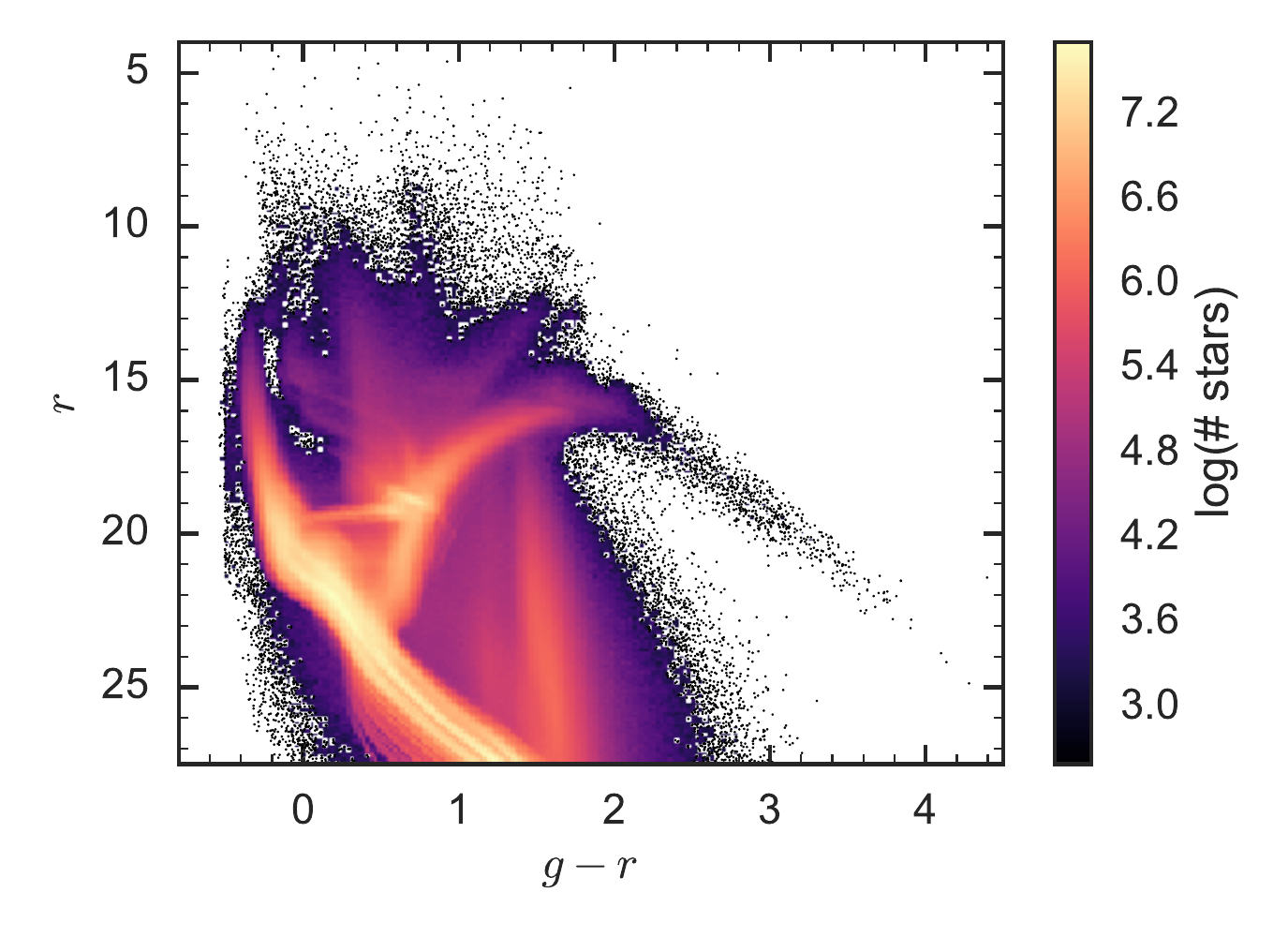}
\includegraphics[width=0.54\textwidth]{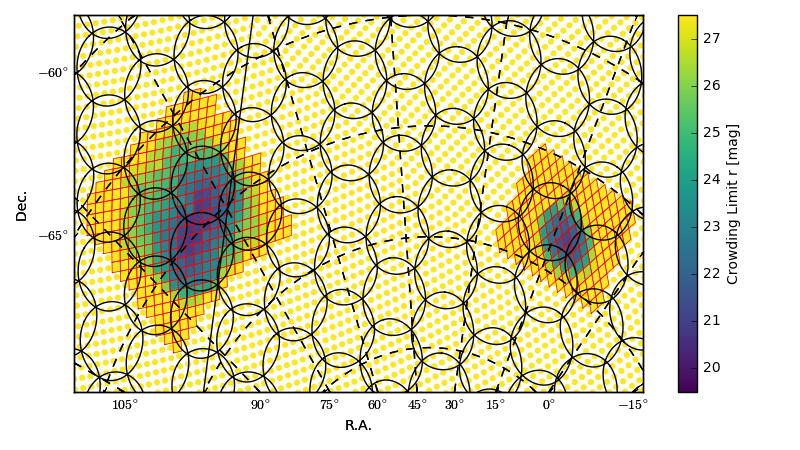}
\caption{Left panel: a CMD illustrating the expected stellar content for the SMC and LMC. It derives from a population synthesis model that uses star-formation history, extinction and distance maps (\cite{Rubele2018} for the SMC, \cite{Harris2009} for the LMC) to extend predictions to deeper magnitudes The MW foreground is included as well \citep[e.g.][]{Girardi2005}. Right panel: Estimated crowding limit, i.e. the $r$ band magnitude that ensures 0.1~mag r.m.s. photometric errors for a seeing of 0.4 arcsec \citep[cf.][]{Olsen2003}, in the central parts of the Magellanic Clouds, compared to the LSST field positions. }
\label{fig:MCs}
\end{figure}

One of the byproducts of the MC main body portion of the survey will be a deep, confusion-limited image of the Clouds, which will be important for establishing the mean luminosity of variable objects and as an LSST legacy data product.  For this, we require that a small number of visits (3 per field for a total of 36 visits) be taken in the best possible seeing conditions, which we estimate to be $\sim$0.4 arcsec, and which will yield a confusion limit 2 mags deeper than under 1 arcsec seeing.  Fig.~\ref{fig:MCs} illustrates the confusion limit in $r$, at which photometric errors become larger than 0.1~mag for a seeing of 0.4~arcsec. Such magnitudes are deep enough to reach the oldest main-sequence turn-offs across most of the MC system, being limited (down to the $\sim4$-Gyr turn-off) only in the densest parts of the LMC and SMC bars.  

\subsection{Individual image depth and/or sky brightness}
For the WFD and SCP portions of the survey, we request 30-second single-snap visits, achieving approximate depths of $(ugrizy)=(23.98, 24.91, 24.42, 23.97, 23.38, 22.47)$.  For the SCP survey, we do not request the $y$ filter.  For the MC main body observations, we request 15-second exposures to limit saturation and to meet the demands of the cadence (see below).

\subsection{Co-added image depth and/or total number of visits}
The visits and depths in the table below are only approximate, and do not include $y$-band, for which we have no strong constraint.  For the detection of dwarf companions and Milky Way/MC structure in the WFD and SCP surveys, the depths are driven by the desire to detect faint main sequence turnoff stars in $gri$ and to use $uz$ for star/galaxy separation and for stellar metallicities.  For the MC main body survey, the visits are driven by the desire to maximize the number of epochs for interstellar scintillation (which will be done in $g$), epochs for variables (done mainly in $gri$) and the static Legacy survey of the Clouds (to be done in all bands to the confusion limit).

\begin{table}[h!]
\begin{tabular}{ |l|c|c|c|c|c|c|c|c|c|c|c|c| }
 \hline
&
 \multicolumn{6}{|c|}{$N_{\rm visits}$}
&
 \multicolumn{6}{|c|}{Depth}\\
 & u & g & r & i & z & y & u & g & r & i & z & y \\
 \hline
 WFD & 165 & 165 & 165 & 165 & 165 & 0 & 26.4 & 27.6  & 27.2 & 26.6 & 25.7 & -- \\
 \hline
 SCP & 40 & 40 & 40 & 40 & 40 & 0 & 25.6 & 26.8 & 26.4 & 25.8 & 24.9 & -- \\
\hline
 MC Main Body & 50 & 1300 & 300 & 300 & 30 & 20 & 
\multicolumn{6}{|c|}{Confusion-limited} \\
 \hline
\end{tabular}
\end{table}


\vspace{1in}
\subsection{Distribution of visits over time}

{\bf WFD and SCP}:
For the WFD and SCP portions of this proposal, the main cadence-dependent goals are the detection of RR Lyrae and the measurement of proper motions.  Both can be accomplished with a homogeneous distribution of visits in time (RR Lyrae, being short-period variables, would benefit from some coverage at short timescales).  For RR Lyrae, the main requirement is 30 visits per filter per field, where the 10 year baseline gives  the advantage to analyze the secondary modulations on a timescale longer than main oscillations (Blazhko phenomenon for RR Lyrae for example).\\
{\bf MC Main Body fields}:
For the 12 MC main body fields: \\
1) The need for rapid timescale observations is driven by the interstellar scintillation experiment, which aims to measure the presence of turbulent interstellar gas clouds down to optical depths of 10$^{-6}$ \cite{Moniez2003}.  The timescale of the scintillation is a few minutes, and is best detected with a campaign of continuous 15-second exposures in $g$.  We will use 1000 total visits in $g$ to measure this signal and to measure short-period variables such as $\delta$ Scuti, SX Phoenicis stars, and white dwarf pulsators.  In order to avoid aliasing on day timescales, we will split the campaign into two groups separated by a few months.\\
2) For periodic variables, we will use 300 visits in each of $gri$ to measure their light curves.  We propose a logarithmically spaced cadence in which we use a quarter of the visits each year, starting with 75 in the first year and finishing with $\sim20$ in the last year.  We have no strong constraint on the concurrence of observations in different filters.
3) Longer timescale transients, such as light echoes ($\sim$weeks) and microlensing from potential stellar mass black holes ($\sim$months) will be served well by the same cadence as the periodic variables.\\

\subsection{Filter choice}
For the SCP mini-survey, we exclude $y$.  We include $y$ in the MC Main Body survey for the purpose of producing a legacy image of the Clouds.

\subsection{Exposure constraints}
For the WFD and SCP surveys, we desire 30-second visits with single snaps.  For the MC Main Body survey, we adopt 15-second exposures in order to sample rapid timescale variability and to limit saturation.
\subsection{Other constraints}
No other constraints.

\subsection{Estimated time requirement}
The survey proposed here relies almost entirely on the reconfiguration of the main survey proposed by the ``Big Sky" (Olsen et al.) white paper, and thus fits within the 90\% time assigned to the main survey.  For the MC Main Body portion of the survey, aimed at observing the Clouds in the time domain, we request 667$\times$12=8004 15-second visits, or 33 hours of exposure time.  These would all be done in the $g$ filter, and would be done in campaign mode, such that overheads will be a small fraction of the total time.

\vspace{.3in}

\begin{table}[ht]
    \centering
    \begin{tabular}{l|l|l|l}
        \toprule
        Properties & Importance \hspace{.3in} \\
        \midrule
        Image quality & 2    \\
        Sky brightness & 2 \\
        Individual image depth & 1  \\
        Co-added image depth & 1  \\
        Number of exposures in a visit   & 1  \\
        Number of visits (in a night)  & 2  \\ 
        Total number of visits & 1  \\
        Time between visits (in a night) & 2 \\
        Time between visits (between nights)  & 1  \\
        Long-term gaps between visits & 1 \\
        Other (please add other constraints as needed) & \\
        \bottomrule
    \end{tabular}
    \caption{{\bf Constraint Rankings:} Summary of the relative importance of various survey strategy constraints, ranked from 1=very important, 2=somewhat important, 3=not important. The main driver for the WFD and SCP portions of this survey are the total number of visits.  For the MC Main Body survey, the measurement of the full variable population depends on the number of visits, the individual exposure time, and the cadence.}
        \label{tab:obs_constraints}
\end{table}

\subsection{Technical trades}
\begin{enumerate}
    \item The basis for this proposal derived from the trade-off between survey area and number of visits explored in the ``Big Sky" (Olsen et al.) white paper.  We used the tools in \url{https://github.com/LSSTScienceCollaborations/survey_strategy_wp} to explore trades between survey area and depth for the Magellanic Clouds survey.  With the reconfigured Big Sky WFD footprint, much more of the southern sky, including the Magellanic Cloud main bodies, falls within WFD than in \ttt{baseline2018a}.  Without this basis, the cost of this proposal would increase dramatically.  
    \item With the exception of the MC Main Body survey, this science is not very sensitive to non-unformity of the cadence, as long as there is sufficient time baseline for measurement of proper motions.  For the MC Main Body survey, requirements are for coverage of all timescales, from $\sim$minutes to 10 years.
    \item For the WFD and SCP areas, we rely on single 30-second snaps to reduce overhead and increase the number of available visits.  Reducing the exposure time for WFD would have significant effect on the number of visits.  For the MC Main Body survey, 15-second exposures are preferred for reduced saturation.
    \item For the WFD and SCP areas, which mainly rely on the static depth, real-time exposure time adjustment could benefit the science.
\end{enumerate}

\section{Performance Evaluation}
The main metric for this proposal is:
\begin{itemize}
    \item The total number of visits with Dec$<60\deg$ in the WFD and SCP areas, and the total number of visits spent on 12 fields covering the Magellanic Cloud main bodies.
\end{itemize}
Further optimization can be obtained by considering the following metrics:
\begin{itemize}
    \item For the detection of Magellanic Cloud structure and dwarf companions in the WFD and SCP areas, a metric is the number of such objects that can be detected.  This number is proportional to area, and depends on depth through the stellar luminosity function for the streams and dwarfs, the foreground contamination, the background galaxy contamination, and the integrated luminosity function of the streams and dwarfs.  One could also account for the ability to use proper motions to remove contamination.  Such a quantitative metric needs further work and testing, which will make use of the simulated LSST catalog developed by co-author Girardi and available through the \href{https://datalab.noao.edu}{NOAO Data Lab}.  To estimate the effect of the $\sim1$mag shallower depth in the SCP region compared to WFD, we can use the stellar luminosity function of an old (10 Gyr), metal-poor ([M/H]=-1.5) population at 50 kpc as a guide.  There are $\sim2\times$ more stars in such a population for $r<27.2$ (WFD) than for $r<26.4$ (SCP), which translates to a factor of $\sqrt{2}$ in S/N of candidate object detections.
\item For the variable population measured in the MC Main Body survey, the specific metrics contained in the WPs by Street et al., Bell
and Lund, Bell and Hermes, and Clementini and Musella will apply. This includes using the UniformityMetric, CampaignLengthMetric, IntervalsBetweenObs, PeriodicStarMetric as well as the aliasing strength MAF
 developed by \cite{Lund2016}.
\item For the Legacy MC Main Body Image, CrowdingMetric applies.  As a rule of thumb, the confusion limit will increase by 0.75 mag for a factor 2 increase in best image quality above 0.4 arcsec.  
\end{itemize}

\vspace{.6in}

\section{Special Data Processing}
The standard LSST pipeline should be sufficient to process the data requested in this proposal.

\section{Acknowledgements}
This work was developed within the Stars, Milky Way, and Local Volume (SMWLV) and Transients and Variable Stars (TVS) Science Collaborations and the authors acknowledge the support of SMWLV and TVS in the preparation of this paper.

\bibliographystyle{plain}
\bibliography{mcs.bib}

\end{document}